\documentclass[letterpaper,twocolumn,prl,aps,superscriptaddress,amsmath,amssymb,floatfix]{revtex4-1}
\usepackage{mathptmx}
\DeclareMathAlphabet{\mathcal}{OMS}{cmsy}{m}{n}
\usepackage[latin9]{inputenc}
\usepackage{color}
\usepackage{verbatim}
\usepackage{float}
\usepackage{amsmath}
\usepackage{amssymb}
\usepackage{graphicx}
\usepackage{esint}
\usepackage[unicode=true,
 bookmarks=true,bookmarksnumbered=false,bookmarksopen=false,
 breaklinks=false,pdfborder={0 0 1},backref=false,colorlinks=true]
 {hyperref}
\hypersetup{
 linkcolor=magenta,urlcolor=blue,citecolor=blue,pdfstartview={FitH},hyperfootnotes=false}

\makeatletter

\usepackage{textcomp}
\usepackage{epstopdf}

\pdfpageheight\paperheight
\pdfpagewidth\paperwidth

\providecommand{\tabularnewline}{\\}

\@ifundefined{textcolor}{}{%
 \definecolor{BLACK}{gray}{0}
 \definecolor{WHITE}{gray}{1}
 \definecolor{RED}{rgb}{1,0,0}
 \definecolor{GREEN}{rgb}{0,1,0}
 \definecolor{BLUE}{rgb}{0,0,1}
 \definecolor{CYAN}{cmyk}{1,0,0,0}
 \definecolor{MAGENTA}{cmyk}{0,1,0,0}
 \definecolor{YELLOW}{cmyk}{0,0,1,0}
}

\usepackage{xcolor}
\usepackage{soul}
\newcommand{\bra}[1]{\ensuremath{\left\langle#1\right|}}
\newcommand{\ket}[1]{\ensuremath{\left|#1\right\rangle}}

\definecolor{blue}{rgb}{0,0,1}
\definecolor{red}{rgb}{1,0,0}
\definecolor{green}{rgb}{0,1,0}

\@ifundefined{textcolor}{}{%
 \definecolor{BLACK}{gray}{0}
 \definecolor{WHITE}{gray}{1}
 \definecolor{RED}{rgb}{1,0,0}
 \definecolor{GREEN}{rgb}{0,1,0}
 \definecolor{BLUE}{rgb}{0,0,1}
 \definecolor{CYAN}{cmyk}{1,0,0,0}
 \definecolor{MAGENTA}{cmyk}{0,1,0,0}
 \definecolor{YELLOW}{cmyk}{0,0,1,0}
}

\usepackage{xcolor}\usepackage{soul}
\newcommand{\overlap}[2]{\left\langle #1 | #2 \right\rangle}
\newcommand{\abs}[1]{\left| #1 \right|}

\definecolor{blue}{rgb}{0,0,1}
\definecolor{red}{rgb}{1,0,0}
\definecolor{green}{rgb}{0,1,0}

\newcommand{\set}[1]{\left\{ #1 \right\}}

\usepackage{soul}

\makeatother

\begin{document}
\title{Unambiguous discrimination of general quantum operations}

\affiliation{Center for Quantum Information, Institute for Interdisciplinary Information
Sciences, Tsinghua University, Beijing 100084, China}
\affiliation{Beijing Academy of Quantum Information Sciences, Beijing 100193, China}
\affiliation{CAS Key Laboratory of Quantum Information, University of Science and Technology of China, Hefei 230026, China}
\affiliation{Hefei National Laboratory, Hefei 230088, China}

\author{Weizhou~Cai}
\affiliation{Center for Quantum Information, Institute for Interdisciplinary Information
Sciences, Tsinghua University, Beijing 100084, China}

\author{Jing-Ning~Zhang}
\email{zhangjn@baqis.ac.cn}
\affiliation{Beijing Academy of Quantum Information Sciences, Beijing 100193, China}

\author{Ziyue~Hua}
\affiliation{Center for Quantum Information, Institute for Interdisciplinary Information
Sciences, Tsinghua University, Beijing 100084, China}

\author{Weiting~Wang}
\affiliation{Center for Quantum Information, Institute for Interdisciplinary Information
Sciences, Tsinghua University, Beijing 100084, China}

\author{Xiaoxuan~Pan}
\affiliation{Center for Quantum Information, Institute for Interdisciplinary Information
Sciences, Tsinghua University, Beijing 100084, China}

\author{Xinyu~Liu}
\affiliation{Center for Quantum Information, Institute for Interdisciplinary Information
Sciences, Tsinghua University, Beijing 100084, China}

\author{Yuwei~Ma}
\affiliation{Center for Quantum Information, Institute for Interdisciplinary Information
Sciences, Tsinghua University, Beijing 100084, China}

\author{Ling~Hu}
\affiliation{Center for Quantum Information, Institute for Interdisciplinary Information
Sciences, Tsinghua University, Beijing 100084, China}

\author{Xianghao~Mu}
\affiliation{Center for Quantum Information, Institute for Interdisciplinary Information
Sciences, Tsinghua University, Beijing 100084, China}

\author{Haiyan~Wang}
\affiliation{Center for Quantum Information, Institute for Interdisciplinary Information
Sciences, Tsinghua University, Beijing 100084, China}

\author{Yipu~Song}
\affiliation{Center for Quantum Information, Institute for Interdisciplinary Information
Sciences, Tsinghua University, Beijing 100084, China}
\affiliation{Hefei National Laboratory, Hefei 230088, China}

\author{Chang-Ling~Zou}
\email{clzou321@ustc.edu.cn}
\affiliation{CAS Key Laboratory of Quantum Information, University of Science and Technology of China, Hefei 230026, China}
\affiliation{Hefei National Laboratory, Hefei 230088, China}

\author{Luyan~Sun}
\email{luyansun@tsinghua.edu.cn}
\affiliation{Center for Quantum Information, Institute for Interdisciplinary Information
Sciences, Tsinghua University, Beijing 100084, China}
\affiliation{Hefei National Laboratory, Hefei 230088, China}


\begin{abstract}
The discrimination of quantum operations has long been an intriguing challenge, with theoretical research~\cite{acin2001statistical,chefles2003retrodiction} significantly advancing our understanding of the quantum features in discriminating quantum objects~\cite{holevo1973statistical, helstrom1976quantum}. 
This challenge is closely related to the discrimination of quantum states, and proof-of-principle demonstrations of the latter have already been realized using optical photons~\cite{mohseni2004optical, becerra2013implementation}.
However, the experimental demonstration of discriminating general quantum operations, including both unitary and non-unitary operations, has remained elusive. In general quantum systems, especially those with high dimensions, the preparation of arbitrary quantum states and the implementation of arbitrary quantum operations and generalized measurements are non-trivial tasks.
Here, for the first time, we experimentally demonstrate the optimal unambiguous discrimination of up to 6 displacement operators and the unambiguous discrimination of non-unitary quantum  operations. 
Our results demonstrate powerful tools for experimental research in quantum information processing~\cite{nielsen2002quantum} and are expected to stimulate a wide range of valuable applications in the field of quantum sensing~\cite{pirandola2018advances}.
\end{abstract}


\maketitle
The discrimination of quantum objects is crucial in quantum mechanics as it enables the retrieval of quantum information by distinguishing and identifying quantum states and operations. This capability forms the foundation for various quantum technologies such as quantum communication~\cite{Bennett1992, Bennett1992a, Weedbrook2012, Izumi2021}, cryptography~\cite{Duifmmodeselsesfiek2000, Gisin2002},  and sensing~\cite{pirandola2018advances}. 
For example, the interactions of the probe light with different mediums constitute quantum operations that are discriminated to detect targets or environmental noises in the field of quantum sensing, such as quantum reading~\cite{Pirandola2011,DallArno2012,Ortolano2021}, illumination~\cite{Lloyd2008,Tan2008,Guha2009,Sanz2017,Qian2023PRL}, and radar~\cite{Assouly2023}. 
Perfect discrimination is preferred for these applications; however, it is only achievable with orthogonal states and is not applicable to general quantum states and operations. Therefore, finding and implementing the optimal strategy to discriminate a set of quantum states or quantum operations presents a fundamental challenge. One of the most appealing strategies, unambiguous discrimination (UD)~\cite{chefles1998unambiguous}, addresses this limitation by allowing for a portion of inconclusive results. 

Over the past decades, the UD has garnered significant attention in both theoretical and experimental researches~\cite{chefles1998unambiguous, Sun2002, Chefles2004, Herzog2005, Wang2006, Chefles2007, Sedlak2009, Kleinmann2010, Waldherr2012, becerra2013implementation, Agnew2014}. Although UD of quantum states has been demonstrated and applied in various scenarios, the exploration of UD of quantum operations is rare~\cite{Laing2009,Zhuang2020a,Zhuang2020b,DeBry2023}. Instead of implementing an optimal detection strategy for discriminating a fixed set of states in the UD of quantum states, the UD of quantum operations requires both the optimal initial probe state and detection strategy. For example, the experimental demonstrations of UD in unitary operations of high-dimensional systems have only been recently achieved in trapped-ion systems~\cite{DeBry2023}. However, for general non-unitary operations, not only are theoretical studies of the UD strategy to saturate the minimal inconclusive outcomes challenging, but experimental demonstrations are also still lacking.


\begin{figure}
\centering{}\includegraphics[width=\columnwidth]{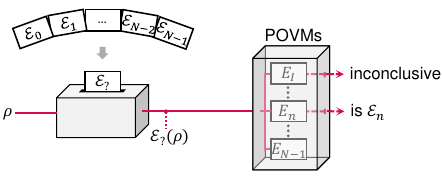}
\caption{\textbf{Unambiguous discrimination of quantum operations.} A quantum black box is able to perform a random quantum operation selected from a given set $\set{{\mathcal{E}}_0,\ldots,{\mathcal{E}}_{N-1}}$. By sending a $d$-dimensional probe state into the black box and performing a POVM measurement on the output states, one can retrodict which operation has been implemented without error, on the cost that some trials lead to inconclusive results.}
\label{fig1}
\end{figure}

\begin{figure*}
\centering{}\includegraphics[width=2\columnwidth]{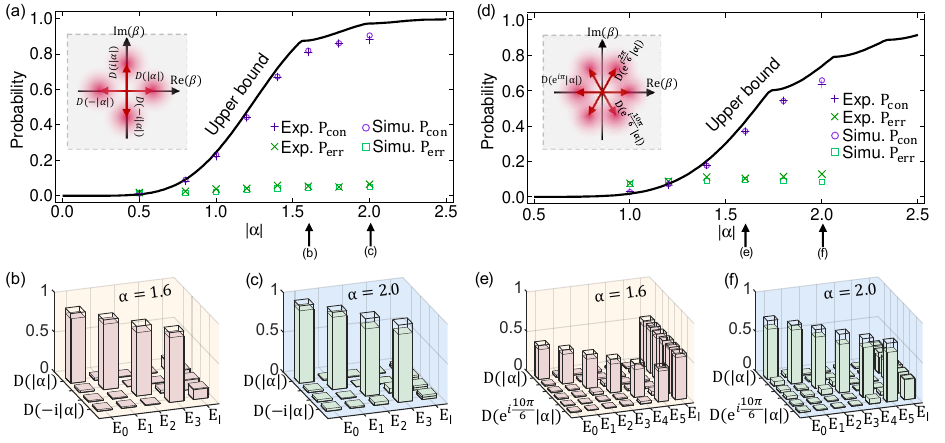}
\caption{\textbf{Unambiguous discrimination of high-dimensional unitary operations.} \textbf{(a)} Conclusive ($P_{\rm con}$) and erroneous ($P_{\rm err}$) probabilities for unambiguously discriminating four-fold displacement operators as functions of the displacement amplitude $\abs{\alpha}$. 
Experimental results for $P_{\rm con}$ and $P_{\rm err}$ are indicated by plus and cross markers, while circles and boxes are the corresponding results that are obtained from numerical simulation considering various experimental decoherence channels. The black solid line shows the analytic results of the optimal conclusive probability. The inset schematically shows the arrangement of the four displacement operators, given the probe state $\ket{\psi_p}=\ket{0}$, with the same amplitude. \textbf{(b, c)} Conditional probabilities of the POVM measurements conditioned on the chosen displacement operators for displacement amplitudes $\abs{\alpha}=1.6$ \textbf{(b)} and $2.0$ \textbf{(c)}. Filled and open bars represent the experimental and numerical results, respectively. The POVM measurement for each operator is repeated 50,000 times to get the output statistics, and the unshown error bars are below the level of $10^{-2}$. \textbf{(d)} The same as \textbf{(a)} for six-fold displacement operators. \text{(e, f)} The same as \text{(b, c)} for six-fold displacement operators. In \textbf{(a)} and \textbf{(d)}, the standard deviations for each set of experimental data points are calculated from repeating the experiments 10 times. These standard deviations are smaller than the markers and are not shown in the figures.}
\label{fig2}
\end{figure*}

Here, we present the first experimental demonstration of UD for general quantum operations in a high-dimensional quantum system, showcasing a protocol for implementing UD in a general experimental setup from a practical perspective.
Firstly, the protocol is experimentally applied to the discrimination of unitary operations,  achieving the optimal UD of up to six displacement operators. These demonstrations present the highest conclusive probabilities compared to previous works~\cite{becerra2013implementation,Izumi2021} in discriminating four displacement operators, and set a record of unambiguously discriminating six displacement operators.
Furthermore, the UD of non-unitary quantum operations is experimentally demonstrated for the first time, specifically for a set of block-dephasing operations and a set of block-Pauli operations in a 4-dimensional quantum system. These demonstrations mark an extension of the discrimination of unitary quantum operations to the discrimination of general quantum operations.
Our results potentially open up new avenues in the experimental research of quantum information processing and stimulate applications in the field of quantum sensing.  

Figure~\ref{fig1} illustrates the schematic of the general framework for UD of quantum operations. It assumes there exists a quantum black box which randomly implements a quantum operation, sampled from a predetermined set of quantum operations $\set{{\mathcal{E}}_n}$ with $n=0,\ldots,N-1$ on a $d$-dimensional quantum system.
A discrimination procedure is as follows: 1) Prepare the principal system to a probe state $\ket{\psi_p}$. 2) Implement a quantum operation ${\mathcal{E}}_n$ to the probe state based on the random choice made by the black box. 3) Perform a generalized measurement~\cite{Zhao2015PRA,Bian2015PRL,Bent2015PRX,Wang2023PRL} such that the label $n$ of the implemented quantum operation can be correctly inferred from the measurement outcome, although with a probability of obtaining inconclusive results. 

Mathematically, the requirements for a probe state $\ket{\psi_p}$ and the detection to achieve UD are
\begin{eqnarray}
{\rm Tr}\left[\hat E_m{\mathcal E}_n\left(\ket{\psi_p}\bra{\psi_p}\right) \right]=P_n\delta_{m,n},~\forall m, n\in\left[0,N-1\right],
\label{eq:requirement_1}
\end{eqnarray}
with the measurement being described by the positive operator-valued measure (POVM)  formalism $\left\{\hat E_I, \hat E_0,\ldots, \hat E_{N-1}\right\}$ for convenience and $0<P_n\leq1$ being the discrimination probability when ${\mathcal E}_n$ has been implemented. The physical constraint of the POVM measurement requires that $\sum_n\hat E_n\leq\hat{\mathbb I}$, such that the operator corresponding to the inconclusive outcome is non-negative, i.e. $\hat E_I\equiv\hat{\mathbb I}-\sum_{n=0}^{N-1}\hat E_n\geq0$. We note that, with a fixed probe state $\ket{\psi_p}$, the discrimination of quantum operations reduces to that of quantum states~\cite{chefles1998unambiguous, feng2004unambiguous}. Moreover, the total conclusive probability $P_{\rm con}=\sum_{n=0}^{N-1}q_nP_n$, with $q_n$ being the probability of occurrence of ${\mathcal E}_n$, can be optimized over the probe state and the POVM measurements. Here we assume the quantum operations are sampled from a uniform probability distribution, i.e. $q_n=1/N$, throughout this paper.  



The following question naturally emerges: Can we determine whether a set of quantum operations can be unambiguously discriminated before looking for the optimal strategies?  
First, the criteria for UD of quantum operations on a quantum system, with the assistance of entangled states between the quantum system and an additionally introduced high-dimensional ancillary system, have been well studied in previous works and can also be found in the Methods.
In the following, we address the above question for more general situations without involving entanglement with ancillary systems (Fig.~\ref{fig1}). We define the union of the supports of all output states as $S\equiv\bigcup_i{\rm supp}\left(\hat\rho_{i}\right)\subseteq {\mathcal H}_S$ and the union of the supports of all output states except for the $n$-th one as $S_n\equiv\bigcup_{i\neq n}{\rm supp}\left(\hat\rho_{i}\right)$ for convenience, where $\rm supp(\hat \rho)$ is the space spanned by the eigenstates of $\hat \rho$ and ${\mathcal H}_S$ is the Hilbert space of the high-dimensional system.
A set of quantum operations cannot be unambiguously discriminated if, for any quantum operation ${\mathcal E}_n$ with $n\in[0,N-1]$, the union of the supports of the other output states always equals that of all output states, i.e. $S_n=S$, for arbitrary pure probe state $\ket{\psi_p}$. 
It is worth noting that we only consider pure states as the probe state, i.e. $\ket{\psi_p}\in{\mathcal H}_S$, while mixed states generally have worse performance for UD. 
Therefore, a practical criterion for unambiguously discriminating a set of quantum operations without entanglement is that the POVM measurement scheme for the UD of general quantum operations $\set{{\mathcal E}_0,\ldots,{\mathcal E}_{N-1}}$ must be able to unambiguously discriminate the output mixed states $\hat\rho_n={\mathcal E}_n\left(\ket{\psi_p}\bra{\psi_p}\right)$, when given a pure probe state $\ket{\psi_p}\in{\mathcal H}_S$. 
As proposed in Ref.~\cite{feng2004unambiguous}, the POVM element $\hat E_n$ that heralds the implementation of ${\mathcal E}_n$ is proportional to the projection operator for the complementary space $S/S_n$. 
By using a probe state $\ket{\psi_p}$, which is described by $2d-1$ independent real variables and represents an arbitrary pure state of a $d$-dimensional system, one can directly determine whether the set of quantum operations can be unambiguously discriminated according to the above criterion by calculating the supports of the set of the output states. 

The experimental demonstrations are carried out with a superconducting circuit device~\cite{hu2019quantum}, which consists of a high-quality microwave cavity and a transmon qubit. The detailed information of the device can be found in the Methods. The microwave mode can be modeled as a quantum harmonic oscillator, with the Hamiltonian given by $\hat H_S=\hbar\omega\hat a^\dag\hat a$, where $\hat a^\dag$ ($\hat a$) is the creation (annihilation) operator and $\omega$ is the transition frequency. An arbitrary displacement operation $\hat D(\alpha)=\exp\left(\alpha\hat a^\dag-\alpha^*\hat a\right)$ can be implemented by directly driving the cavity with a resonant microwave pulse. Meanwhile, the low-lying $d$ eigenstates of the cavity can be used to encode a quantum $d$-dimensional system or a qudit. Under this scenario, universal manipulations, including preparation of both arbitrary states and unitary operations, can be achieved by utilizing a dispersively coupled transmon qubit. Moreover, arbitrary quantum operations can be efficiently implemented by real-time feedback controls, as proposed and experimentally demonstrated in Refs.~\cite{shen2017quantum} and \cite{Cai2021}.
Therefore, all the experimental steps for UD of quantum operations can be implemented on the same device without extra transportation or conversion of quantum states, thus minimizing the experimental decoherence in the protocol. 

To begin with, we examine the discrimination of a set of $N$ displacement operators $\hat D\left(\alpha_n\right)$ with $\alpha_n=|\alpha|\exp\left(\frac{i2n\pi}{N}\right)$ with $n=0, \ldots, N-1$. 
A natural choice for the probe state is the ground state $\ket{0}$ of the cavity, which makes the task of discriminating the displacement operators equivalent to discriminating linearly independent symmetric states, i.e. $\ket{\psi_n}=\hat D\left(\alpha_n\right)\ket{0}$. Previous theoretical investigations~\cite{chefles1998unambiguous, chefles1998optimum} have provided the expression for the optimal UD probability. However, the existing experiments~\cite{becerra2013implementation, Izumi2021} have not yet saturated the upper bound. 

Our UD protocol can saturate the optimal discrimination probability and is constructed as follows. We first construct a set of states $\ket{\psi_n^{\perp}}$, by the Gram-Schmidt orthogonalization procedure (see the Methods),  such that $\left|\overlap{\psi_n^\perp}{\psi_{n'}}\right|\propto \delta_{n,n'}$ for all $n,n'=0,1,\ldots,N-1$. 
To satisfy the requirement in Eq.~(\ref{eq:requirement_1}), the operator $\hat E_n$ is proportional to the projector of $\ket{\psi_n^\perp}$, i.e. $\hat E_n=\frac{P}{\abs{\overlap{\psi_n}{\psi_n^\perp}}^2}\ket{\psi_n^\perp}\bra{\psi_n^\perp}$, where the conditional discrimination probability $P\in(0,1)$ can be optimized to make the POVM measurements physically valid. 
It is straightforward to verify that this discrimination scheme is unambiguous in the ideal case, with the total conclusive and inconclusive probabilities being $P_{\rm con}=P$ and $P_{\rm inc}=1-P$, respectively. As derived in  Ref.~\cite{chefles1998optimum}, the upper bound for the conclusive probability is given by $P_{\mathrm{con}} \leq N\times{\rm min}_r\left|c_r\right|^2$, where $\left|c_r\right|^2=N^{-2}\sum_{n,n'}\exp\left[-i2\pi r\left(n-n'\right)/N\right]\overlap{\psi_n}{\psi_{n'}}$ with $r=0,1,\ldots,N-1$.
In experiments, however, there will be a non-vanishing probability of obtaining erroneous results, $ P_{\rm err}=1- P_{\rm con}- P_{\rm inc}$, resulted from decoherence and operation imperfections.

We experimentally implement the optimal POVM measurement scheme and saturate the upper bound of UD for the first time. The results for the UD of $N=4$ and $N=6$ displacement operators with various values of the displacement amplitude $\abs{\alpha}$ (up to $\abs{\alpha}=2$) are shown in Fig.~\ref{fig2}.  
The insets of  Figs.~\ref{fig2}(a) and \ref{fig2}(d) schematically illustrate the positions of the coherent states in the phase space, where the overlaps between the coherent states decrease as $\abs{\alpha}$ increases. Consequently, the ideal conclusive probability $P_{\rm con}$ monotonically converges to unity, as indicated by the solid lines in Figs.~\ref{fig2}(a) and \ref{fig2}(d). 
The experimental results of the conclusive probability $P_{\rm con}$ are in good agreement with numerical simulations that properly consider the amplitude damping of the cavity and the dephasing and relaxation of the transmon qubit. Figures~\ref{fig2}(b, c) and \ref{fig2}(e, f) show the conclusive, inconclusive, and error probabilities for each displacement operator. 
The numerically estimated error budget of the performance can be found in the Methods. The discrepancy between the experiment and the theoretical upper bound is primarily attributed to the decoherence of the auxiliary qubit.

The experiment shown in Fig.~\ref{fig2} demonstrates that a set of non-orthogonal coherent states, generated by a set of displacement operators, can be unambiguously discriminated. These coherent states are commonly used in long-distance transmission tasks, where they can carry information with different phases, such as in phase shift keying protocols~\cite{Izumi2021,Kanitschar2022}. Additionally, increasing the number of displacement operators that are used to generate coherent states allows the transmission of more information. However, the non-orthogonality of these coherent states, deteriorated by amplitude damping noises during transmission, limits the transmission distance and leads to ambiguous information. Our experiment demonstrates the successful retrieval of information from these non-orthogonal coherent states using the UD strategy.

\begin{figure*}
\centering{}\includegraphics[width=2\columnwidth]{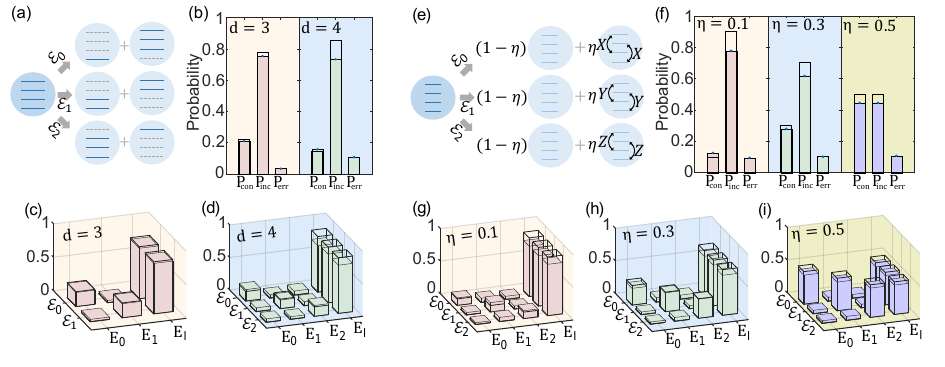}
\caption{\textbf{Unambiguous discrimination of high-dimensional non-unitary operations.} \textbf{(a)} Schematic representation of the block-dephasing operations in the 4-dimensional Hilbert space. \textbf{(b)} Conclusive ($P_{\rm con}$), inconclusive ($P_{\rm inc}$), and erroneous ($P_{\rm err}$) probabilities for the UD of block-dephasing operations in $d$-dimensional Hilbert spaces with $d = 3$ (left) and $4$ (right). Filled and open bars represent the experimental and numerical results, respectively. \textbf{(c)} Output probability distribution of the POVM measurement conditioned on the chosen block-dephasing operations in a $3$-dimensional Hilbert space. \textbf{(d)} The same as \textbf{(c)} for $d=4$. \textbf{(e)} Schematic representation of the identity operations with block-Pauli errors. \textbf{(f)} Conclusive, inconclusive, and erroneous probabilities for the UD of block-Pauli operations in a $4$-dimensional Hilbert space with the block-Pauli error rate $\eta = 0.1$ (left), $0.3$ (middle), and $0.5$ (right).  \textbf{(g-i)} Output probability distributions of the POVM measurements conditioned on the block-Pauli errors that have occurred with the error rate $\eta = 0.1$ \textbf{(g)}, $0.3$ \textbf{(h)}, and $0.5$ \textbf{(i)}.}
\label{fig3}
\end{figure*}

Having shown the ability to implement arbitrary quantum operations in our experimental system~\cite{Cai2021}, we extend the UD of unitary operations to a more general and complex situation, i.e. the UD of high-dimensional non-unitary quantum operations.
Here, we use the low-lying four states of the cavity, denoted by $\ket{0}$, $\ket{1}$, $\ket{2}$, and $\ket{3}$,  to encode a quantum $4$-dimensional system and implement a set of three block-dephasing operations $\set{{\mathcal E}_0, {\mathcal E}_1, {\mathcal E}_2}$. These operations are defined as follows: ${\mathcal E}_0\left(\hat\rho\right)={\mathcal P}_{\set{0}}\left(\hat\rho\right)+{\mathcal P}_{\set{123}}\left(\hat\rho\right)$, ${\mathcal E}_1\left(\hat\rho\right)={\mathcal P}_{\set{01}}\left(\hat\rho\right)+{\mathcal P}_{\set{23}}\left(\hat\rho\right)$, and ${\mathcal E}_2\left(\hat\rho\right)={\mathcal P}_{\set{012}}\left(\hat\rho\right)+{\mathcal P}_{\set{3}}\left(\hat\rho\right)$, as schematically shown in Fig.~\ref{fig3}(a). Here, ${\mathcal P}_{s}=\sum_{i,j\in s}\ket{i}\bra{i}\hat\rho\ket{j}\bra{j}$ denotes the projector into the subspace spanned by the set of states labeled by $s$. We choose the probe state as the equally-populated superposition state $\ket{\psi_p}=\frac{1}{2}\left(\ket{0}+\ket{1}+\ket{2}+\ket{3}\right)$. With this choice, the POVM measurement exists (our choice of the POVM can be found in the Methods). In addition, we also demonstrate a set of two block-dephasing operations $\{{\mathcal E}_0, {\mathcal E}_1\}$ on the first three dimensions of the quantum system, where ${\mathcal E}_0 \left(\hat\rho\right)={\mathcal P}_{\set{01}}\left(\hat\rho\right)+{\mathcal P}_{\set{2}}\left(\hat\rho\right)$, ${\mathcal E}_1 \left(\hat\rho\right)={\mathcal P}_{\set{0}}\left(\hat\rho\right)+{\mathcal P}_{\set{12}}\left(\hat\rho\right)$, and probe state is $\ket{\psi_p}=\frac{1}{\sqrt{3}}\left(\ket{0}+\ket{1}+\ket{2}\right)$. The measurement results of the conclusive, inconclusive, and erroneous probabilities are shown in Fig.~\ref{fig3}(b) left (for $d=3$) and right (for $d=4$). We measure the conditional probability $P\left(m|n,\psi_p\right)$ and present the results in Figs.~\ref{fig3}(c) for 3-dimensional and \ref{fig3}(d) for 4-dimensional. Here, the conditional probability $P\left(m|n,\psi_p\right)$ represents the probability of obtaining the outcome $m$ corresponding to $E_m$ when given the probe state $\ket{\psi_p}$ and the operation ${\mathcal E}_n$. In comparison to the ideal case, we can clearly observe a portion of erroneous results emerging due to experimental imperfections, as shown in Figs.~\ref{fig3}(b-d). 

To demonstrate the flexibility of our setup, we showcase the UD of a set of parameterized quantum operations. As shown in Fig.~\ref{fig3}(e), these quantum operations are represented in the operator-sum form as ${\mathcal E}_n\left(\hat\rho,\eta\right)=(1-\eta)\hat K_{n,0}\hat\rho\hat K_{n,0}^\dag+\eta\hat K_{n,1}\hat\rho\hat K_{n,1}$, with $n\in\set{0, 1, 2}$. Here, the Kraus operators are $\hat K_{n,0}=\hat{\mathbb I}_4$, $\hat K_{0,1}=\hat{X}_{\{0,2\}}+\hat{X}_{\{1,3\}}$, $\hat K_{1,1}=\hat{Y}_{\{0,2\}}+\hat{Y}_{\{1,3\}}$, and $\hat K_{2,1}=\hat{Z}_{\{0,2\}}+\hat{Z}_{\{1,3\}}$, with $\hat{X}$, $\hat{Y}$, and $\hat{Z}$ being the Pauli operators in the 2-dimensional subspace spanned by the basis states specified in the subscripts. Intuitively, the identity part within these operations contributes to the inconclusive probability $(1-\eta)$. 
We select the state $\ket{\psi_p}=\frac{1}{\sqrt{2}}\left(\ket{0}+\ket{3}\right)$ as the probe state. With this choice, the POVM measurement exists (our choice of the POVM can be found in the Methods). The measurement results for the conclusive, inconclusive, and erroneous probabilities are presented in Fig.~\ref{fig3}(f). Additionally, Figs.~\ref{fig3}(g-i) display the measured conditional probabilities $P\left(m|n,\psi_p\right)$ for three different parameter values: $\eta = 0.1$, $0.3$, and $0.5$. The error budget of this experiment with  $\eta = 0.5$, obtained through simulation, can be found in the Methods. 

The application of UD to non-unitary quantum operations provides an efficient method to classify and characterize open quantum processes or quantum noises in a single use. 
Hence, one potential application scenario of UD is when transient coherent or incoherent interactions impact a quantum system and researchers can only distinguish the possible quantum processes in a single use. 

In conclusion, the UD of general quantum operations is experimentally demonstrated. With complete control over a high-dimensional system of a bosonic mode to realize POVM measurements, we experimentally demonstrate the UD of four and six displacement operations, achieving the highest conclusive probability to date. In principle, our experimental protocol can allow the conclusive probability to saturate the theoretical upper bound~\cite{chefles1998optimum}. Furthermore, we extend the UD to non-unitary quantum operations, both block-dephasing operations and block-Pauli operations, in a high-dimensional system. The bosonic modes equipped with arbitrary quantum operations~\cite{Cai2021,Han2021} provide a unique testbed for verifying quantum protocols of quantum Shannon theory, exploiting fundamental physics regarding quantum information transmission over quantum channels~\cite{Wilde2013,Watrous2018}, and extending these insights to more practical applications in quantum sensing~\cite{Lloyd2008,Tan2008,Guha2009,Pirandola2011,DallArno2012,Sanz2017,pirandola2018advances,Assouly2023}. 
Our experiments also represent a shift of quantum information processing from closed quantum systems towards open quantum systems. In the foreseeable future, we anticipate the implementation of more quantum information technologies in open quantum systems.


\smallskip{}



%

\vbox{}
\renewcommand{\figurename}{Extended Data FIG.}
\renewcommand{\tablename}{Extended Data TABLE}
\setcounter{figure}{0}

\begin{figure*}[hbt]
	\includegraphics{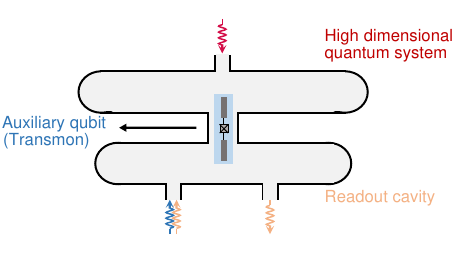} \caption{\textbf{Schematic diagram of the experimental device.} }
	\label{fig:S_device}
\end{figure*}

\clearpage{}
\noindent \textbf{\large{}Methods}{\large\par}
\noindent\textbf{Experimental setup}.
Our experimental device, as shown in Extended Data FIG.~\ref{fig:S_device}, consists of a superconducting transmon qubit that is dispersively coupled to two rectangular microwave cavities, both machined from high-purity 5N5 aluminum. 
One of these cavities serves as a high-dimensional quantum system with a frequency of $\omega_{s}/2\pi=7.634\,~\mathrm{GHz}$ and a singe photon lifetime of 143~$\mu s$. The other cavity, referred to as the readout cavity, operates at a frequency of $\omega_{r}/2\pi=8.610\,~\mathrm{GHz}$ and has a short lifetime of 44~ns for fast readout of the transmon qubit. The transmon qubit, fabricated on a sapphire substrate with a frequency of $\omega_{a}/2\pi=5.692\,~\mathrm{GHz}$, a relaxation time of $30~\mu$s, and a pure dephasing time of $120~\mu$s, is used as an auxiliary qubit to assist the high-dimensional quantum system in implementing arbitrary quantum operations. The dispersive coupling strengths between the transmon qubit and the high-dimensional quantum system, as well as the readout cavity, are $\chi_\mathrm{qs}/2\pi=1.90\,~\mathrm{MHz}$ and $\chi_\mathrm{qr}/2\pi=3.65\,~\mathrm{MHz}$, respectively. 

Our setup incorporates a Josephson parametric amplifier that enables high-fidelity quantum non-demolition (QND) single-shot measurements of the auxiliary qubit, with a measurement time of 320~ns. These measurements yield a fidelity of 99.9\% for $\ket{g}$ and 98.9\% for $\ket{e}$ of the auxiliary qubit. The infidelity in the QND measurements is mainly attributable to the decoherence of the auxiliary qubit. High-fidelity QND single-shot measurements enable real-time adaptive control, a crucial technology for realizing arbitrary quantum operations~\cite{lloyd2001engineering, shen2017quantum, Cai2021}. The adaptive control selects and implements unitary gates on both the high-dimensional quantum system and the auxiliary transmon qubit based on the states of the auxiliary qubit. This functionality is achieved by using field programmable gate arrays with home-made logics. Additional details about our experimental device and setup can be found in Refs.~\cite{hu2019quantum, Cai2021}. 

\vbox{}

\noindent\textbf{Construction of reciprocal states.}
Given a set of linearly independent states, denoted as $\{ \ket{\psi_0},\ldots,\ket{\psi_{N-1}}\}$, the reciprocal states $\ket{\psi_n^\perp}$ are defined to be the states satisfying $\overlap{\psi_n^\perp}{\psi_{n'}}\propto \delta_{n,n'}$ for all $n,n'=0,1,\ldots,N-1$. These states are important for the construction of the POVM that is capable of performing UD of the given set of states. To find out the explicit form of the reciprocal states, we need to perform $N$ times of the Gram-Schmidt orthogonalization. Specifically, to obtain $\ket{\psi_n^\perp}$, we first obtain a set of orthonormal basis $\left\{\ket{u_m^{(n)}}\right\}_{m=0}^{N-2}$, which spans the $(N-1)$-dimensional Hilbert space spanned by all $\ket{\psi_{n'}}$ with $n'\neq n$. Then the reciprocal state $\ket{\psi_n^\perp}$ can be obtained as being proportional to the remaining of $\ket{\psi_n}$ after subtracting the components parallel to all $\ket{u_m^{(n)}}$.  Mathematically, $\ket{\psi_n^\perp}$ can be written as
\begin{eqnarray}
\ket{\psi_n^\perp}=\ket{\psi_n}-\sum_{m=0}^{N-2}\left\langle u_m^{(n)} \Big| \psi_n \right\rangle\ket{u_m^{(n)}},
\end{eqnarray}
up to normalization. This procedure is repeated for $N$ times to obtain the whole set of the reciprocal states.

\vbox{}

\noindent\textbf{Experimental realization of arbitrary quantum operations and positive operator-valued measures (POVMs).} 
An arbitrary quantum operation $\mathcal{E}$ with a Kraus rank $N$ can be represented by a set of Kraus operators $\{K_0, K_1,...,  K_{N-1}\}$, where $\mathcal{E}(\rho)=\sum_i K_i\rho K_i^{\dagger}$. For instance, a quantum operation $\mathcal{E}$ with Kraus rank 4 has Kraus operators $\{K_0, K_1, K_2, K_3\}$. To implement this quantum operation, a two-layer binary-tree quantum circuit needs to be constructed. In the first layer, we implement the Kraus rank-2 quantum operation with Kraus operators $\{A_0,A_1\}$, where

\begin{equation}
\begin{aligned}
A_{0}=\sqrt{K_{0}^{\dagger}K_0+K_{1}^{\dagger}K_1}, \\
A_{1}=\sqrt{K_{2}^{\dagger}K_2+K_{3}^{\dagger}K_3}.
\end{aligned}
\label{eq:channelLayer1}
\end{equation}

After the implementation of the first layer, we perform a projective measurement on the auxiliary qubit with respect to the states $\{\ket{0}\bra{0},\ket{1}\bra{1}\}$. The outcomes $\{0,1\}$ correspond to the implemented Kraus operators $\{A_0,A_1\}$, respectively. In the second layer, we implement four Kraus operators $\{B_{00},B_{01},B_{10},B_{11}\}$ based on the measurement results of the auxiliary qubit in the first layer. When the measurement result is $0$, we implement Kraus operators $B_{00}$ and $B_{01}$; otherwise, we implement $B_{10}$ and $ B_{11}$. The Kraus operators $\{B_{00},B_{01},B_{10},B_{11}\}$ are given by: 
\begin{equation}
\begin{aligned}
B_{00}=K_0A_0^{-1}, \\
B_{01}=K_1A_0^{-1}, \\
B_{10}=K_2A_1^{-1}, \\
B_{11}=K_3A_1^{-1}.
\end{aligned}
\label{eq:channelLayer2}
\end{equation}

The measurement results of the auxiliary qubit in the two layers have outcomes $\{00, 01, 10, 11\}$ corresponding to the implemented Kraus operators $\{K_0, K_1, K_2, K_3\}$, respectively. The rank-2 quantum operation in each layer is realized by joint unitary gates between the quantum system and the auxiliary qubit with the form:
\begin{equation}
\begin{aligned}
U_{A}=\left(\begin{array}{ll}
A_0 & * \\
A_1 & * 
\end{array}\right), 
U_{B0}=\left(\begin{array}{ll}
B_{00} & * \\
B_{01} & * 
\end{array}\right), 
U_{B1}=\left(\begin{array}{ll}
B_{10} & * \\
B_{11} & * 
\end{array}\right).
\end{aligned}
\end{equation}
Here the symbol ``*" denotes that the corresponding elements in the unitary matrices do not contribute to the quantum operation $\mathcal{E}$, but should satisfy $U_{A}^{\dagger}U_{A}=U_{B0}^{\dagger}U_{B0}=U_{B1}^{\dagger}U_{B1}=I$.

For a quantum system of a cavity that is dispersively coupled to a qubit, arbitrary unitary gates can be realized by using selective number-dependent arbitrary phase gates~\cite{Krastanov2015} or the gradient ascent pulse engineering (GRAPE) method~\cite{Khaneja2005,deFouquieres2011,Heeres2017}. In our work, the unitary gates in the two-layer binary tree are realized by using the GRAPE method.

The realization of POVMs is similar to the realization of arbitrary quantum operations. For example, a POVM with four elements is described by $\{E_I,E_0,E_1,E_2\}$ with $\sum_i E_i =I$. Since all elements are Hermitian and semidefinite, we can construct a set of Kraus operators $\{e_I, e_0, e_1, e_2\}$ with
\begin{equation}
\begin{aligned}
e_{I}=\sqrt{E_I}, \\
e_{0}=\sqrt{E_0}, \\
e_{1}=\sqrt{E_1}, \\
e_{2}=\sqrt{E_2}.
\end{aligned}
\label{eq:POVM_Kraus}
\end{equation}
The measurement result of each element on a density matrix $\rho$ can be represented by
\begin{equation}
\begin{aligned}
\mathrm{Tr}[E_I\rho]=\mathrm{Tr}[e_I\rho e_I^{\dagger}], \\
\mathrm{Tr}[E_0\rho]=\mathrm{Tr}[e_0\rho e_0^{\dagger}], \\
\mathrm{Tr}[E_1\rho]=\mathrm{Tr}[e_1\rho e_1^{\dagger}], \\
\mathrm{Tr}[E_2\rho]=\mathrm{Tr}[e_2\rho e_2^{\dagger}].
\end{aligned}
\label{eq:POVM_results}
\end{equation}
Therefore, the implementation of this POVM is the same as the implementation of a quantum operation with four Kraus operators $\{e_I, e_0, e_1, e_2\}$, which can be realized by the two-layer binary-tree method. By recording the measurement outcomes of the auxiliary qubit in each layer, we can obtain the measurement results of each POVM element. The probabilities of the measurement results of the qubit $\{P_{00},P_{01},P_{10},P_{11}\}$ correspond to the measurement of the POVM elements $\{E_I,E_0,E_1,E_2\}$, respectively.

More details of the method to construct and implement arbitrary quantum operations and arbitrary POVMs on high-dimensional quantum systems can also be found in Ref.~\cite{Cai2021}.

\vbox{}

\noindent\textbf{POVM for discriminating block-dephasing operations.}
For 3-dimensional block-dephasing operations, the POVM measurements are given by
$$
E_I=I-E_0-E_1,
$$
$$
E_0=\left(\begin{array}{lll}
1 & -1 & 0 \\
-1 & 1 & 0 \\
  0 & 0 & 0
\end{array}\right)/3,
$$
$$
E_1=\left(\begin{array}{lll}
0 & 0 & 0 \\
0 & 1 & -1 \\
0 & -1 & 1
\end{array}\right)/3.
$$

For 4-dimensional block-dephasing operations, the POVM measurements are given by
$$
E_I=I-E_0-E_1-E_2,
$$
$$
E_0=\left(\begin{array}{llll}
1 & -1 & 0 & 0 \\
-1 & 1 & 0 & 0 \\
0 & 0 & 0 & 0 \\
0 & 0 & 0 & 0
\end{array}\right)/(2+\sqrt{2}),
$$
$$
E_1=\left(\begin{array}{llll}
0 & 0 & 0 & 0 \\
0 & 1 & -1 & 0 \\
0 & -1 & 1 & 0 \\
0 & 0 & 0 & 0
\end{array}\right)/(2+\sqrt{2}),
$$
$$
E_2=\left(\begin{array}{llll}
0 & 0 & 0 & 0 \\
0 & 0 & 0 & 0 \\
0 & 0 & 1 & -1 \\
0 & 0 & -1 & 1
\end{array}\right)/(2+\sqrt{2}).
$$

\vbox{}

\noindent\textbf{POVM for discriminating block-Pauli operations.} The POVM measurements are given by
$$
E_I=I-E_0-E_1-E_2,
$$
$$
E_0=\left(\begin{array}{llll}
1 & 0 & 0 & -1 \\
0 & 0 & 0 & 0 \\
0 & 0 & 0 & 0 \\
-1 & 0 & 0 & 1
\end{array}\right)/2,
$$
$$
E_1=\left(\begin{array}{llll}
0 & 0 & 0 & 0 \\
0 & 1 & 1 & 0 \\
0 & 1 & 1 & 0 \\
0 & 0 & 0 & 0
\end{array}\right)/2,
$$
$$
E_2=\left(\begin{array}{llll}
0 & 0 & 0 & 0 \\
0 & 1 & -1 & 0 \\
0 & -1 & 1 & 0 \\
0 & 0 & 0 & 0
\end{array}\right)/2.
$$

\vbox{}

\noindent\textbf{Error Budget}.
To analyze the error source in our experiments, we numerically simulate the experimental processes using the same pulse sequences that are employed in the actual experiments. To quantify the performance of POVMs, we introduce a distance defined as $\mathcal{D}=\sum|A_\mathrm{simu}-A_\mathrm{ideal}|/N$. Here $\sum$ represents a summation over all matrix elements, $A_\mathrm{simu}$ and $A_\mathrm{ideal}$ denote the results obtained from simulated and ideal POVMs, respectively, and $N$ is the total number of the prepared quantum operations. We simulate the process and calculate the distance for the unambiguous discrimination (UD) of four-fold displacement operators with $|\alpha|=1.6$ under various conditions. These conditions include processes with both auxiliary qubit decoherence and system decoherence (resulting in $\mathcal{D}=0.116$), without auxiliary qubit decoherence but with system decoherence ($\mathcal{D}=0.050$), and without any auxiliary qubit or system decoherence ($\mathcal{D}=0.021$). The main errors for the case without any auxiliary qubit or system decoherence are attributed to imperfections in the pulse sequence. These imperfections include two types of errors. The first type is optimization errors in the pulse optimization process with the GRAPE method. The second type is the resetting error of the auxiliary qubit, which is induced by a large dispersive shift due to the high photon numbers in the quantum system. The distances under all conditions are presented in Table~\ref{Table:coherent_UD}.

To quantify the performance of the probe state preparation, implementation of quantum operations, and POVMs, we simulate the process for the UD of block-Pauli operations in a 4-dimensional Hilbert space with a block-Pauli error rate $\eta=0.5$. We also calculate the state fidelities of the prepared probe states with auxiliary qubit decoherence and system decoherence ($\mathcal{F}_s=98.7\%$), without auxiliary qubit decoherence but with system decoherence ($\mathcal{F}_s=99.5\%$), and without any auxiliary qubit or system decoherence ($\mathcal{F}_s=100\%$). 
The state fidelity is computed using the formula $\mathcal{F}_s=\mathrm{Tr}(\rho_\mathrm{simu}\rho_\mathrm{ideal})$, where $\rho_\mathrm{ideal}$ is the density matrix of the ideal pure state and $\rho_\mathrm{simu}$ is the density matrix of the simulated probe state. We then calculate the process fidelities of the implemented quantum operations with auxiliary qubit decoherence and system decoherence ($\mathcal{F}_P=95.4\%$ for block-Pauli X, $\mathcal{F}_P=95.7\%$ for block-Pauli Y, $\mathcal{F}_P=96.2\%$ for block-Pauli Z), without auxiliary qubit decoherence but with system decoherence ($\mathcal{F}_P=97.9\%$ for block-Pauli X, $\mathcal{F}_P=98.0\%$ for block-Pauli Y, $\mathcal{F}_P=98.1\%$ for block-Pauli Z), and without any auxiliary qutbit or system decoherence ($\mathcal{F}_P=99.7\%$ for block-Pauli X, $\mathcal{F}_P=99.6\%$ for block-Pauli Y, $\mathcal{F}_P=99.9\%$ for block-Pauli Z). All of these quantum operations are implemented on an ideal probe state. 
The process fidelity is computed using the formula $\mathcal{F}_P=\mathrm{Tr}(\sqrt{\sqrt{\chi_\mathrm{ideal}}\chi_\mathrm{simu}\sqrt{\chi_\mathrm{ideal}}})^2$, where $\chi_\mathrm{ideal}$ is the ideal process matrix and $\chi_\mathrm{simu}$ is the simulated process matrix. 
We also calculate the distances of the POVMs with auxiliary qubit decoherence and system decoherence ($\mathcal{D}=0.100$), without auxiliary qubit decoherence but with system decoherence ($\mathcal{D}=0.037$), and without any auxiliary qubit or system decoherence ($\mathcal{D}=0.009$). The distances of POVMs are calculated based on ideal probe states and ideal implementation of quantum operations.
The results of the state fidelities, process fidelities, and distances are summarized in Table~\ref{Table:block_pauli_UD}.

\vbox{}

\noindent\textbf{Criteria for unambiguous discrimination.}
Here we review previous theoretical research on the discrimination of general quantum operations, including both unitary operations and non-unitary operations. An elegant criterion for a set of unitary gates that can be unambiguously discriminated is that if and only if the set of unitary operations are linearly independent~\cite{chefles2003retrodiction,Chefles2007}. Without entangled probe states, such a condition reduces to the local linear dependence for the set of unitary operations~\cite{chefles2003retrodiction}.
For non-unitary operations, Ref.~\cite{Wang2006} has proposed a theorem on the UD of quantum operations: a set of quantum operations can be unambiguously discriminated if and only if for any operation in the set, its support is not included in the supports of the rest operations. The definition of the supports of operations can be found in Ref.~\cite{Wang2006}.
However, the derivation, as well as the linearly independent condition for unitary operations, requires the probe state to possess full Schmidt rank, residing in the Hilbert space of a combined system consisting of the quantum system and an ancillary system with a dimension at least as large as the quantum system. In other words, the quantum system and the ancillary system must be initialized in a highly entangled state. Yet, the ability to deterministically generate quantum system-ancilla entanglement, especially for high-dimensional quantum systems, is still challenging. Therefore, we believe that the theoretical study of UD of quantum operations without entanglement is of practical importance. 

With a given probe state, the discrimination of quantum operations reduces to that of quantum states. Thus we briefly review previous theoretical results about the UD of quantum states. It is pointed out in Ref.~\cite{chefles1998unambiguous} that a set of pure quantum states can be unambiguously discriminated if and only if they form a linearly independent set. Later, Ref.~\cite{feng2004unambiguous} proves that the necessary and sufficient condition for UD of a set of mixed quantum states is that the support of an arbitrary state in the set is not included in the supports of the rest states. Therefore, by iterating through all states of the quantum system as the probe state, we can determine whether a set of quantum operations can be unambiguously discriminated according to the criteria for UD of quantum states.
In practice, we describe the probe state with $2d-1$ independent real variables. Such a probe state can represent an arbitrary pure state of the $d$-dimensional quantum system. If the set of output states, which is obtained by implementing the set of operations on the probe state, can be unambiguously discriminated according to the criteria for UD of quantum states, the set of quantum operations can be unambiguously discriminated without entanglement.

\clearpage

\smallskip{}

\noindent \textbf{\large{}Data availability}{\large\par}

\noindent All data generated or analysed during this study are available
within the paper and its Methods. Further source
data will be made available on reasonable request.

\smallskip{}

\noindent \textbf{\large{}Code availability}{\large\par}

\noindent The code used to solve the equations presented in the Methods will be made available on reasonable request.

\smallskip{}

\noindent \textbf{\large{}Acknowledgment}{\large\par}
\noindent This work was supported by the National Natural Science Foundation of China (Grants No.~92165209, 11925404, 92365301, 12204052, and 92265210), Innovation Program for Quantum Science and Technology (No.~2021ZD0300203), Natural Science Foundation of Beijing (Grant No.~Z190012), Fundamental Research Funds for the Central Universities, China Postdoctoral Science Foundation (Grants No. 2023M733409 No. BX2021167), and the National Key Research and Development Program of
China (Grant No. 2017YFA0304303). This work was partially carried out at the USTC Center for Micro and Nanoscale Research and Fabrication.

\smallskip{}
\noindent \textbf{\large{}Author contributions}{\large\par}
\noindent WC performed the experiment and analyzed the data with the assistance of ZH, WW, XP, XL, and YM. LS directed the experiment. CLZ proposed the experiment, JNZ and CLZ provided theoretical support. LH developed the FPGA logic. WC fabricated the JPA. LH and XM fabricated the devices with the assistance of HW, and YPS. WC, JNZ, CLZ, and LS wrote the manuscript with feedback from all authors.

\smallskip{}

\noindent \textbf{\large{}Competing interests}{\large\par}

\noindent The authors declare no competing interests.

\smallskip{}




\begin{table*}[t]
	\vspace{9bp}
	\begin{centering}
		\begin{tabular*}{0.6\textwidth}{@{\extracolsep{\fill}}@{\extracolsep{\fill}}@{\extracolsep{\fill}}@{\extracolsep{\fill}}ccccccc}
			\hline
			\hline					  					
		      error type & error source & distance D (contribution) & total
            \tabularnewline
            \hline
			Decoherence  & auxiliary qubit & 0.066 (57.4$\%$) & 0.095 (81.9$\%$) \tabularnewline
							   & quantum system   & 0.029 (24.5$\%$) \tabularnewline
			\hline
			Control & pulse imperfection  &  0.021 (18.1$\%$) &0.021 (18.1$\%$) \tabularnewline
			\hline
			\hline
		\end{tabular*}
		\par\end{centering}
	\caption{\textbf{Distance $\mathcal{D}$ of the simulated POVM results of the UD of four-fold displacement operators with $|\alpha|=1.6$.} The contribution of each error source is shown in bracket.}
	\label{Table:coherent_UD}
\end{table*}

\begin{table*}[t]
	\vspace{12bp}
	\begin{centering}
		\begin{tabular*}{1\textwidth}{@{\extracolsep{\fill}}@{\extracolsep{\fill}}@{\extracolsep{\fill}}@{\extracolsep{\fill}}ccccccc}
			\hline
			\hline
			error type & error source  & state preparation infidelity  &  X process infidelity &  Y process infidelity &  Z process infidelity & POVMs distance \tabularnewline
			\hline					  					
			
			Decoherence  &auxiliary qubit  & 0.008 (61.7$\%$) &0.025 (54.8$\%$)&0.023 (53.7$\%$)&0.019 (48.9$\%$)&0.063 (63.3$\%$)  \tabularnewline
						 & system  & 0.005 (38.1$\%$) &0.018 (38.2$\%$)&0.016 (37.7$\%$)&0.018 (47.8$\%$)&0.028 (27.7$\%$)  \tabularnewline
			\hline
			Control 	 & pulse imperfection  &  0.000 (0.2$\%$)&0.003 (7.0$\%$)&0.004 (8.6$\%$)&0.001 (3.2$\%$)&0.009 (9.0$\%$)  \tabularnewline
			\hline
			\hline
		\end{tabular*}
		\par\end{centering}
	\caption{\textbf{State infidelities, process infidelities, and distance of the simulated processes of the UD of block-Pauli operations in 4-dimensional Hilbert space with a Pauli rate $\eta=0.5$.} Process infidelities of the block-Pauli operations are labeled by X, Y, and Z, respectively. The contribution of each error source is shown in bracket.}
	\label{Table:block_pauli_UD}
\end{table*}

\end{document}